\documentclass[doublecol,figures]{epl2}
\usepackage{amsmath}
\usepackage{amsfonts}
\usepackage[T1]{fontenc}
\usepackage[english]{babel}
\newcommand{\ie}{\emph{i.e., }}
\newcommand{\eg}{\emph{e.g., }}
\newcommand{\reff}[1]{(\ref{#1})}
\newcommand{\eref}[1]{Eq.\reff{#1}}
\newcommand{\erefs}[1]{Eqs.\reff{#1}}
\newcommand{\figref}[1]{Fig.\ref{#1}}
\newcommand{\citeref}[1]{Ref.\!\cite{#1}}
\newcommand{\citerefs}[1]{Refs.\!\cite{#1}}

\usepackage{color}

\title{Reanalysis of the beam-plasma instability using\\ the Dyson-like equation formalism}
\shorttitle{Beam-plasma system and Dyson-like formalism}

\author{Nakia Carlevaro\inst{1,2} \and Francesco Finelli\inst{3} \and Giovanni Montani\inst{1,4}}
\shortauthor{N. Carlevaro, F. Finelli, G. Montani}

\institute{
\inst{1} ENEA, Fusion and Nuclear Safety Department, C. R. Frascati,
              Via E. Fermi 45, 00044 Frascati (Roma), Italy.\\
\inst{2} Consorzio RFX, Corso Stati Uniti 4, 35127 Padova, Italy.\\
\inst{3} Physics Department ``E. Fermi'', University of Pisa, L.go Bruno Pontecorvo 3, 56127 Pisa, Italy.\\
\inst{4} Physics Department, ``Sapienza'' University of Rome,
              P.le Aldo Moro 5, 00185 Roma, Italy.
}


\abstract{We analyze the problem of the beam-plasma instability via the analytical treatment of the so-called Dyson equation. We first compared the prediction of the model constructed by fixing the electric field amplitude with respect to a $N$-body Hamiltonian numerical simulation. Then, we demonstrate that the shortcomings of such an analytical formulation must be essentially identified with the breaking-down of the self-consistent evolution of the field and the particle distribution function.}

\begin{document}

\maketitle

\section{Introduction}
One of the most interesting paradigm of plasma physics is the so-called beam-plasma system \cite{OM68,OWM71,ZCrmp}. In fact, beside the possible laboratory applications (especially in plasma accelerators \cite{Li14,ES96,KJ86}), this scenario has important conceptual implications in the properties of the radial transport in Tokamak devices \cite{BB90a,nceps16}.

The beam-plasma interaction faces the influence that a tenuous electron beam has on the Langmuir spectrum of a thermalized plasma, pumping up to saturation the resonant modes, \ie those modes whose phase velocity is close to beam particle speed. In the Vlasov-Poisson scheme, the considered theoretical framework is commonly called bump-on-tail (BoT) paradigm \cite{BS11,shalaby17,pommo17} and it was proposed as a toy model for the radial transport in Tokamak experiments \cite{BB90a,BB90b,BB90c,BS11}: the velocity gradients, which trigger the inverse Landau damping, are mapped into the radial pressure gradients of fast ions in a toroidal plasma configuration \cite{GG12,spb16}.

Such a parallelism and the highly non-linear character of the involved physics, make the study of the beam-plasma instability still an actual problem, essentially in order to shed light on the different mechanisms (and their relative relevance) responsible for the observed behavior of energetic particles. In this respect, in \citeref{ncentropy} it has been shown that the beam-palsma dynamics can be properly characterized by the so-called quasi-linear model \cite{LP81,BB95b,BB96b,Laval99} only in the late phase of the evolution, while the temporal mesoscales are characterized by a significant degree of convection in the velocity space (for specific considerations on the role of the self-consistent evolution, see \citeref{mcc18}).

Although the most successful analysis of the beam-plasma interaction has been provided in \citeref{OWM71}, a very interesting and general theoretical framework for the problem has been introduced in \citeref{AK66}. The Vlasov-Poisson equation is addressed via an expansion of the particle distribution function in a power series of the electric field intensity. Then a hierarchy in the different contributions (poles in the Laplace expansion) is determined, by introducing a diagrammatic approach. Moreover, an analytical approach to the solution of the system is derived in terms of an expansion in Hermite polynomials, for the beam-plasma interaction in the presence of a monochromatic field when the saturated amplitude is assumed as constant.

The main aim of the present work is to investigate the predictivity of the analytical treatment presented in \citeref{AK66}, by means of a comparison with a pure numerical $N$-body analysis of the beam-plasma system using the Hamiltonian formulation discussed in \citerefs{CFMZJPP,ncentropy} (and refs. therein), for the monochromatic case.

We clearly demonstrated that the analytical solution presented in \citeref{AK66} fails in predicting the detailed features of the distribution function of the fast electrons interacting with the saturated spectrum. In fact, a significant bumpiness arises in the distribution profile, together with an inversion of the velocity gradient that is not observed in the real simulation experiment. Then, we face the question concerning the nature of such a discrepancy, in principle attributed to the truncation of the expansion in Hermite polynomials. To this end, it is analyzed a Vlasov equation, obtained by suitably recombining the $0^{th}$ and $k^{th}$ components of the Fourier harmonics, \ie the so-called Dyson equation \cite{ZCrmp}. Substituting in this equation a constant amplitude mode, it is shown that the emerging distribution function has the same irregular behavior of the one obtained by the expansion truncation. This analysis clarified that the shortcomings of the analytical solution of the Vlasov equation can not be attributed to mathematical approximation, but they concern the nature of the addressed assumption, \eg the exact constant value of the field amplitude, on which the analysis is built up.

Finally, we insert in the Dyson equation the exact electric field as extracted from the numerical simulation experiment, based on the $N$-body code. The obtained distribution function closely resembles the one produced by the simulations, underlining that the assumption of a saturated constant electric field is a weak hypothesis and it is the source of un-predictivity of the study in \citeref{AK66}. 

This final result has a deep physical meaning, because it clearly shows how the self-consistency of the Vlasov-Poisson evolution is a basic feature of the BoT paradigm: its breaking-down can lead to make a relevant discrepancy on the late time prediction for the system dynamics. For instance, the valuable determination of the overlap of nearly living resonance has been reached by using dynamical system renormalization methods \cite{EE18}, for which the field amplitude is frozen in. Analogously, all the quasi-linear approximation of the transport features of fast ions in a Tokamak are based on assigned spectrum properties, non self-consistently evolved with the same distribution function. Although such transport analyses are well grounded in terms of reliably simplifying assumptions on the spectrum morphology, the study here developed suggests the necessity of a careful evaluation of the coupled field and particle evolution.

The main merit of this investigation consists in stressing how the breaking-down of self-consistency is allowed when we are interested in qualitative features of the transport. If we desire to be able to reproduce the fast ion redistribution, we need to keep the field and particle dynamics strictly coherent.

\section{Vlasov-Poisson equations towards a Dyson formulation}
We start reviewing the main steps of \citeref{AK66}, which lead to a Dyson equation for the 1D beam dynamics. The electron distribution $f(t,x,v)$ and the electric field $\mathcal{E}(t,x)$ are Fourier and Laplace transformed providing
\begin{align}\nonumber
F_k(\omega,v)=\int_0^\infty\!\!\!\!\!f_k(t,v)e^{i\omega t}dt\;,\quad
E_k(\omega)=\int_0^\infty\!\!\!\!\!\mathcal{E}_k(t)e^{i\omega t}dt\;,
\end{align}
respectively, where $f_k(t,v)$ and $\mathcal{E}_k(t)$ denote standard Fourier $k$-components. In this scheme, the Vlasov equation reads as follows:
\begin{align}
F_k(\omega,&v)=\frac{if_k(t=0,v)}{\omega-kv}+\nonumber\\
&+\frac{ie}{m}\sum_{k'+k''=k}\int\frac{d\omega'}{2\pi}\frac{E_{k'}(\omega')}{\omega-kv}
\partial_vF_{k''}(\omega-\omega',v)\;,\label{eq.trans.vlasov}
\end{align}
which is naturally coupled to the Poisson equation reading
\begin{align}\label{eqpoi}
ikE_k(\omega)=-4\pi en_0\int dv F_k(\omega,v)\;,
\end{align}
where $n_0$ is the homogeneous electron density. For each Fourier component of the transformed distribution function the following formal expansion in powers of the electric field is considered
\begin{equation}
\label{eq.formal.expansion}
F_k(\omega,v)=\sum_{n\geq 0}F_k^{(n)}(\omega,v)\,,\quad\text{with}\quad F_k^{(n)}\sim E_k^n\,.
\end{equation}
\begin{figure}[ht]
\centering
\includegraphics[width=0.5\columnwidth]{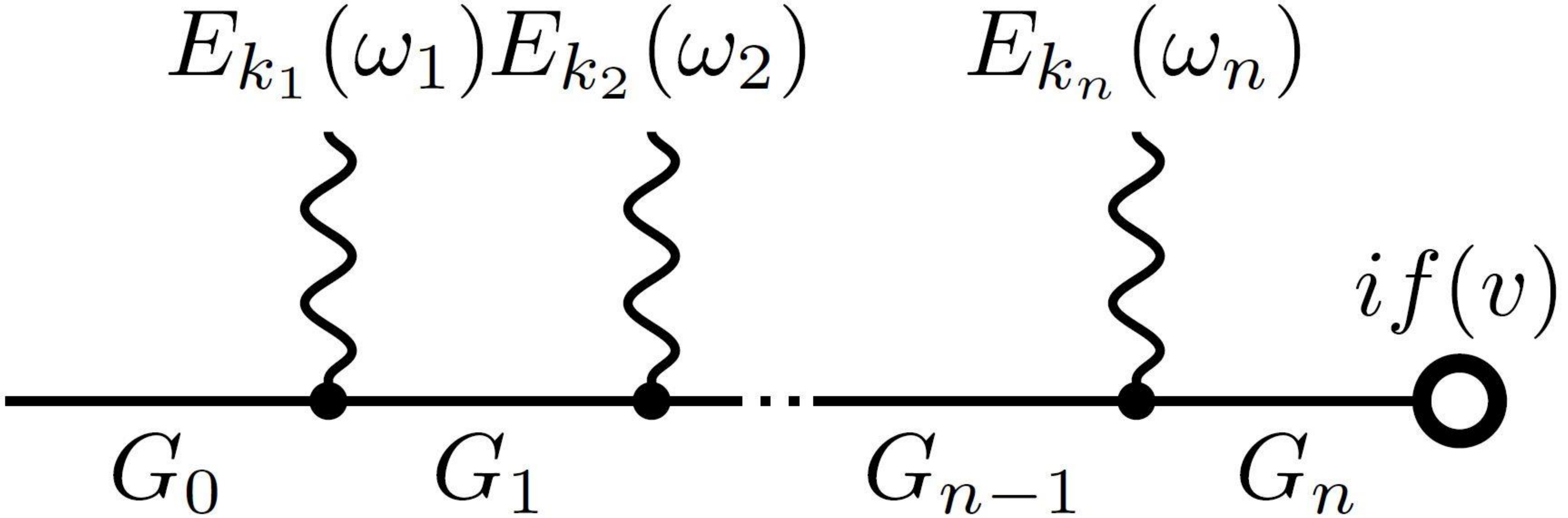}
\caption{Diagrammatic representation of \eref{eq.closed.recursion}.}
\label{fig.closed.recursion.diag}
\end{figure}

Let $f(v)$ be the spatially homogeneous initial distribution, then it is possible to close an expression for the generic $F_k^{(n)}$ via recursion, starting from
\begin{equation}
\label{eq.start.recursion}
F_k^{(0)}(\omega,v)=if(v)\omega^{-1}\delta_{k,0}
\end{equation}
and getting \eref{eq.closed.recursion} (see next page), where $e$ and $m$ are the electron charge and mass, respectively.
\begin{widetext}
\begin{multline}
\label{eq.closed.recursion}
F_k^{(n)}(\omega,v)=\left(\frac{ie}{m}\right)^n\sum_{k_1+\cdots+k_n=k}\int\frac{d\omega_1}{2\pi}\cdots\frac{d\omega_n}{2\pi}\frac{E_{k_1}(\omega_1)}{\omega-kv}\partial_v\frac{E_{k_2}(\omega_2)}{\omega-kv-(\omega_1-k_1v)}\partial_v\cdots\\\cdots\frac{E_{k_n}(\omega_n)}{\omega-kv-\sum_{s=1}^{n-1}(\omega_s-k_sv)}\partial_v\frac{if(v)}{\omega-kv-\sum_{s=1}^n(\omega_s-k_sv)}\,.
\end{multline}
\end{widetext}

In order to deal with this expression, it is convenient to adopt a diagrammatic representation as shown in \figref{fig.closed.recursion.diag}. Here, we defined the propagators
\begin{equation}
\label{eq.propagator}
G_s(\omega,v,\{\omega_j,k_j\})=\big(\omega-kv-\sum_{j=1}^{s}(\omega_j-k_jv)\big)^{-1}
\end{equation}
and the $s$-th interaction vertex represents the operator $(ie/m)\int(d\omega_s/2\pi)\partial_v$. Of course, it is implied a summation over all $k_j$ under the requirement that they sum up to $k$.

In the limiting case in which the electrostatic (plasma) Langmuir modes have constant amplitude, \ie $\mathcal{E}_k(t)=E_k^{(0)}e^{-i\omega_k^Rt}$, with $\omega_k^R$ a real constant (no real frequency shift will be considered), Laplace-transformed modes are readily given by
\begin{equation}
\label{eq.linear.limit}
E_k(\omega)=iE_k^{(0)}\big(\omega-\omega_k^R\big)^{-1}\,,
\end{equation}
thus one can integrate over $\omega_s$ simply replacing it with $\omega_{k_s}^R$. Referring to \figref{fig.closed.recursion.diag}, it is clear that sometimes two adjacent field lines can be one the conjugate of the other, namely $E_k$ and $E_{-k}$, and we will represent it by closing the two lines in a loop as if the same mode were emitted and then absorbed. Because $\omega_{-k}^R=-\omega_k^R$, every time this configuration occurs the propagators external to the loop are identical, meaning that, in order to Laplace anti-transform, one has to integrate at least one non simple pole. At each order $n$ the highest order for a pole is (the integer part of) $n/2+1$, leading to a secularity $t^{n/2}$: this is the reason because we refer to the expansion \eqref{eq.formal.expansion} as formal (no term can be neglected). This is also true in the more general case in which $\omega_k=\omega_k^R+i\gamma_k$ (the real part still constant in time) and the condition $\gamma_k\ll\omega_P$ holds, where $\omega_P$ is the plasma frequency (in order to actually have propagating waves): secularities $t^{n/2}$ are replaced by exponentially growing factors $(\omega_k^R/\gamma_k)^{n/2}$.

\begin{figure}[ht]
\centering
\includegraphics[width=0.5\columnwidth]{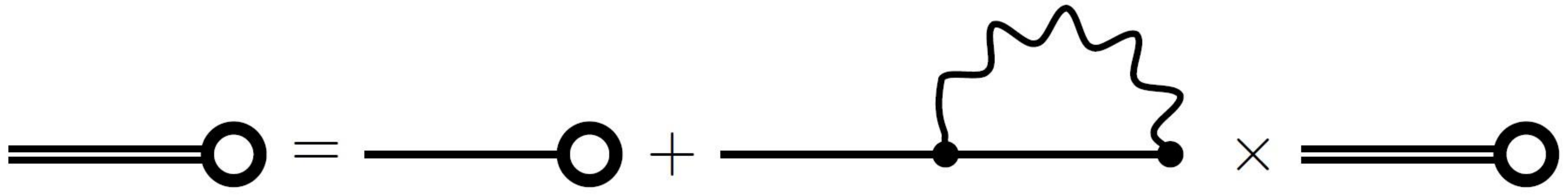}
\caption{Diagrammatic representation of \eref{eq.dyson}.}
\label{fig.dyson.diag}
\end{figure}

Taking a partial resummation of all and only the \emph{heaviest} diagram of each order ($\sim(\omega_k^R/\gamma_k)^{n/2}$), one finds the Dyson equation \cite{Dyson49,ZCrmp}
\begin{align}
F_0(\omega,v)&=\frac{if(v)}{\omega}+\left(\frac{ie}{m}\right)^2\sum_{q=-\infty}^{+\infty}\int\frac{d\omega'}{2\pi}\frac{d\omega''}{2\pi}\frac{E_{-q}(\omega')}{\omega}\times\nonumber\\
&\times\partial_v\frac{E_q(\omega'')}{\omega-\omega'-qv}\partial_vF_0(\omega-\omega^{'}-\omega'',v)\;,\label{eq.dyson}
\end{align}
diagrammatically shown in \figref{fig.dyson.diag}.
It sholud be noted that already in \cite{M64} the diagram thecnique was pioneered in the study of turbulence in plasma physics, obtaining the Dyson equation shown above. In \eref{eq.dyson}, it is worth focusing our attention on the term $F_0$ because it is the most relevant one, being the initial condition itself homogeneous, and because it is easy to show how every $F_k$ can be obtained simply by $F_0$ \cite{AK66}.

For completeness, it should be noted that in \citeref{AK66} this model is closed with the Poisson equation \reff{eqpoi} rewritten in the following form
\begin{subequations}
\begin{align}
\label{eq.poisson}
\int\frac{d\omega'}{2\pi}\epsilon_k(\omega,\omega')E_k(\omega')=-\frac{4\pi e n_0}{k}\int\!\!d v\,\frac{g_k(v)}{\omega-kv}\,,\\
\label{eq.perm.nonlinear}
\epsilon_k(\omega,\omega')=\frac{i}{\omega-\omega'}+\frac{\omega_P^2}{k}\int\!\!\!\frac{d v}{\omega-kv}\partial_vF_0(\omega-\omega',v)\,,
\end{align}
\end{subequations}
where $g_k(v)$ stands for initial spatial inhomogeneities that we consider already of the same order of $F_k^{(1)}$ (\emph{small} compared to $f(v)$). Furthermore, $\epsilon_k$ is a dielectric function of the plasma. In \citeref{AK66}, it is also shown how \erefs{eq.dyson} and \reff{eq.poisson} can be considered as a generalized quasi-linear model \cite{LP81,BB95b,BB96b,Laval99}.

\subsection{A solution for monochromatic field}
Following \citeref{AK66}, let us now investigate an analytic solution for \eref{eq.dyson} in the case of a single Langmuir mode of constant amplitude $E_k^{(0)}$. In the original part of our work, we will outline the physical content and the shortcomings of such an approximation. Starting from \eref{eq.dyson}, we assume the validity of \eref{eq.linear.limit} and we limit the sum over $q=\pm k$, corresponding to neglect harmonics of the fundamental mode as they are higher-order effects \cite{OWM71,L72}, thus getting
\begin{equation}
\label{eq.dyson.mono}
\hat{F}_0(\omega,\xi)=\frac{i\hat{f}(\xi)}{\omega}-\alpha^2\partial_{\xi}\frac{1}{\omega^2-\alpha^2\xi^2}\partial_{\xi}\hat{F}_0(\omega,\xi)\,,
\end{equation}
where we have defined defined
\begin{align}\label{alphapar}
\alpha^2=\sqrt{2}ek|E_k^{(0)}|/m\,,
\end{align}
(this parameter differs from the bounce (trapping) frequency $\omega_B$ for an $\mathcal{O}(1)$ factor) and we have switched the velocity variable from $v$ to $\xi=(k/\alpha)(v-\omega_k^R/k)$, introducing $\hat{f}(\xi)=f(v(\xi))$ and the same for $\hat{F}_0(\omega,\xi)$.

Defining $\Psi(\omega,\xi)=\alpha^2(\omega^2-\alpha^2\xi^2)^{-1}\partial_{\xi}\hat{F}_0(\omega,\xi)$, \eref{eq.dyson.mono} takes the following form
\begin{equation}
\label{eq.dyson.mono.psi}
\partial_{\xi}^2\Psi(\omega,\xi)+\big(\omega^2 \alpha^{-2}-\xi^2\big)\Psi(\omega,\xi)=\frac{i}{\omega}\partial_{\xi}\hat{f}(\xi)
\end{equation}
closely resembling the equation defining the parabolic cylinder functions (PCFs) $\psi_n(\xi)$:
\begin{subequations}
\begin{align}
\label{eq.pcf.def}
\partial_{\xi}^2\psi_n(\xi)+(2n+1-\xi^2)\psi_n(\xi)=0\,,\\
\label{eq.pcf}
\psi_n(\xi)=\frac{e^{-\xi^2/2}}{\sqrt{2^nn!\sqrt{\pi}}}H_n(\xi)\,.
\end{align}
\end{subequations}
Since PCFs are an orthonormal basis for differentiable functions ($H_n(\xi)$ are the Hermite polynomials), one can solve \eref{eq.dyson.mono.psi} by projecting $\Psi(\omega,\xi)$ and $\partial_{\xi}\hat{f}(\xi)$ onto it. The result, written back in time domain, is
\begin{subequations}
\label{eq.ak.result}
\begin{multline}
\label{eq.mono}
\hat{f}_0(t,\xi)=\hat{f}(\xi)+\\
+\sum_{n\geq 0}\frac{\beta_n}{2n+1}\partial_{\xi}\psi_n(\xi)[1-\cos(\alpha\sqrt{2n+1}\,t)]\,,
\end{multline}
with
\begin{equation}
\label{eq.beta}
\beta_n=\int d\xi\,\psi_n(\xi)\partial_{\xi}\hat{f}(\xi)\,.
\end{equation}
\end{subequations}

We conclude by noting how $\alpha$ determines both the time scale of the process ($\sim\alpha^{-1}$), and the non linear velocity spread of the resonance ($\sim\alpha/k$).

\section{Comparison to numerical simulations}
In this Section, we compare the solution in \eref{eq.mono} with respect to the self-consistent dynamics of the beam-plasma system via $N$-body simulations. The aim of this study is to clarify the predictivity of the analytic solution of the Dyson approach presented above versus the real non-linear features of the beam-plasma interaction.

\subsection{Hamiltonian description of the beam-plasma system}
In \citerefs{OM68,OWM71}, the beam-plasma system is modelled as a fast electron beam injected into a 1D plasma. Such a background plasma is considered as a cold linear dielectric medium (a periodic slab of length $L$) which supports longitudinal electrostatic Langmuir waves. This scheme is isomorphic to the well-known BoT paradigm \cite{BGK57,BS11,Br11,shalaby17,pommo17}. The density $n_B$ of the addressed beam is taken much smaller than the density $n_0$ of the background electron plasma, and we introduce the density parameter $\eta$ as $\eta=n_B/n_0$ (of course $\eta\ll 1$). The Langmuir potential $\varphi(x,t)$ is expressed in terms of $M$ Fourier components $\varphi_{k_j}(t)$ with frequency $\omega_j\simeq\omega_P$ for $j=1,\,...,\,M$.

In this work, we use the standard Hamiltonian formulation of the BoT paradigm described in \citerefs{CFMZJPP,ncentropy} (and refs therein), where the broad energetic beam self-consistently evolves in the presence of the set of Langmuir modes taken at the plasma frequency. With this assumption, the cold background dielectric function, \ie $\epsilon=1-\omega_P^2/\omega^2$, results to be nearly vanishing \cite{OM68}. Thus, the Poisson equation for plasma oscillations can be cast as an evolutive equation, and a given mode is linearly unstable if the resonance condition $k_j=\omega_P/v_{rj}$ (where $v_{rj}$ is a selected initial velocity of beam particles) is satisfied. Finally, the force equation describes particle trajectories.

Particle positions along the $x$ direction are labeled by $x_i$ and $N$ denotes the total particle number ($i=1,\,...,\,N$). The beam-plasma system is now governed by the following $N$-body system:
\begin{subequations}
\label{eq.hamiltonian.sys}
\begin{align}
\label{eq.hamiltonian.vel}
&\dot{x}_i(t)=v_i(t)\,,\\
\label{eq.hamiltonian.acc}
&\dot{v}_i(t)=\frac{ie}{m}\sum_{j=1}^M k_j\varphi_{k_j}(t)e^{ik_jx_i(t)}+c.c.\,,\\
\label{eq.hamiltonian.phi}
&\dot{\varphi}_{k_j}(t)=-i\omega_P\varphi_{k_j}(t)+i\eta\frac{4\pi n_0\omega_P}{2k_j^2N}\sum_{i=1}^{N}e^{-ik_jx_i(t)}\,,
\end{align}
\end{subequations}
where the dot represents the time derivative. For the sake of convenience, in the numerical analysis we use the normalization: $\bar{x}_i=x_i(2\pi/L)$, $\tau=t\omega_P$, $u_i=\partial_\tau\bar{x}_i=v_i(2\pi/L)/\omega_P$, $\ell_j=k_j(2\pi/L)^{-1}$, $\phi_j=(2\pi/L)^2 e\varphi_{k_j}/m\omega_P^2$. The resonance conditions now rewrite $\ell_j u_{rj}=1$, with $\ell_j$ taken as an integer number (best approximation of $1/u_r$).

\begin{figure}[ht]
\centering
\includegraphics[width=0.8\columnwidth]{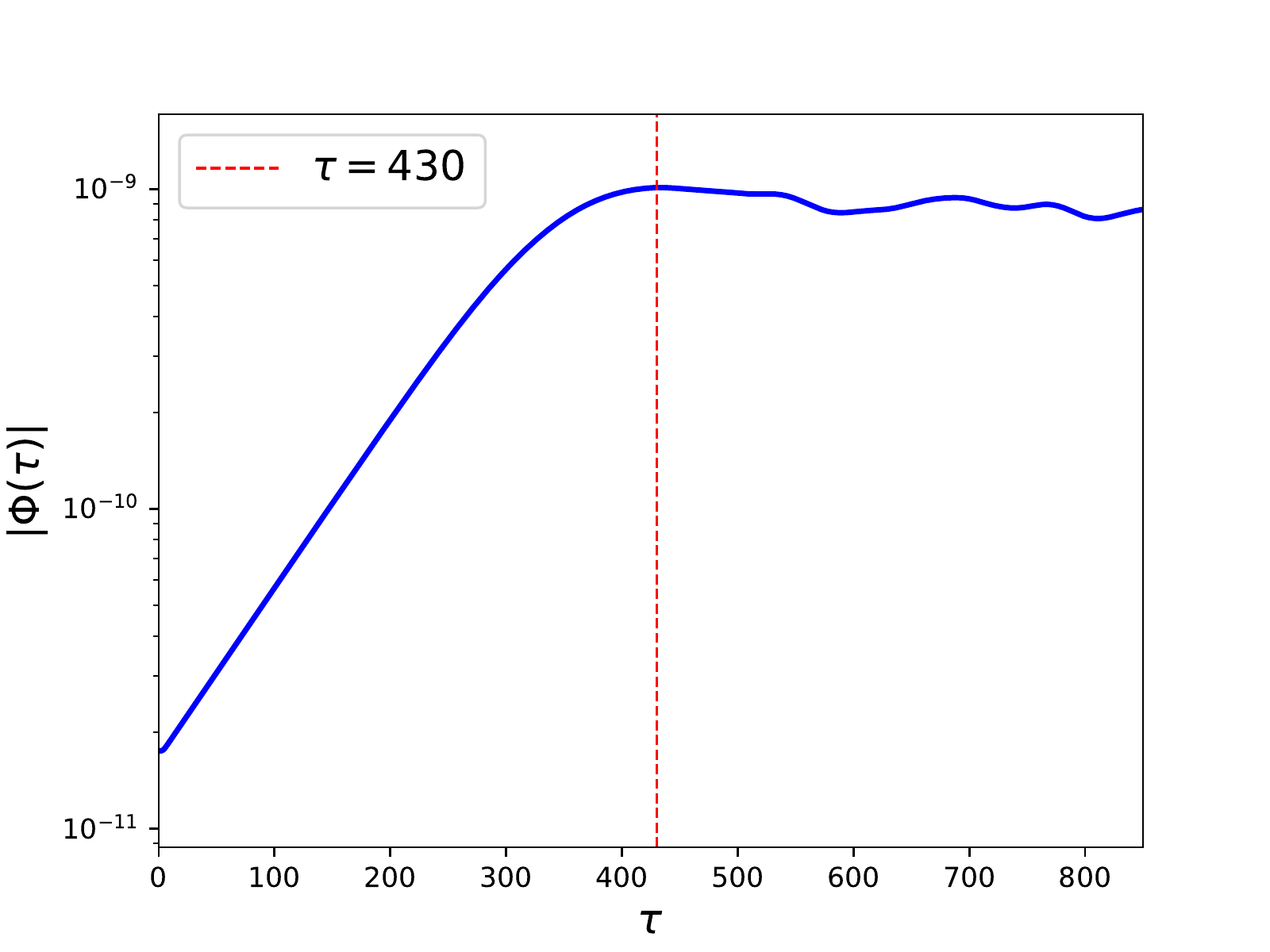}
\includegraphics[width=0.8\columnwidth]{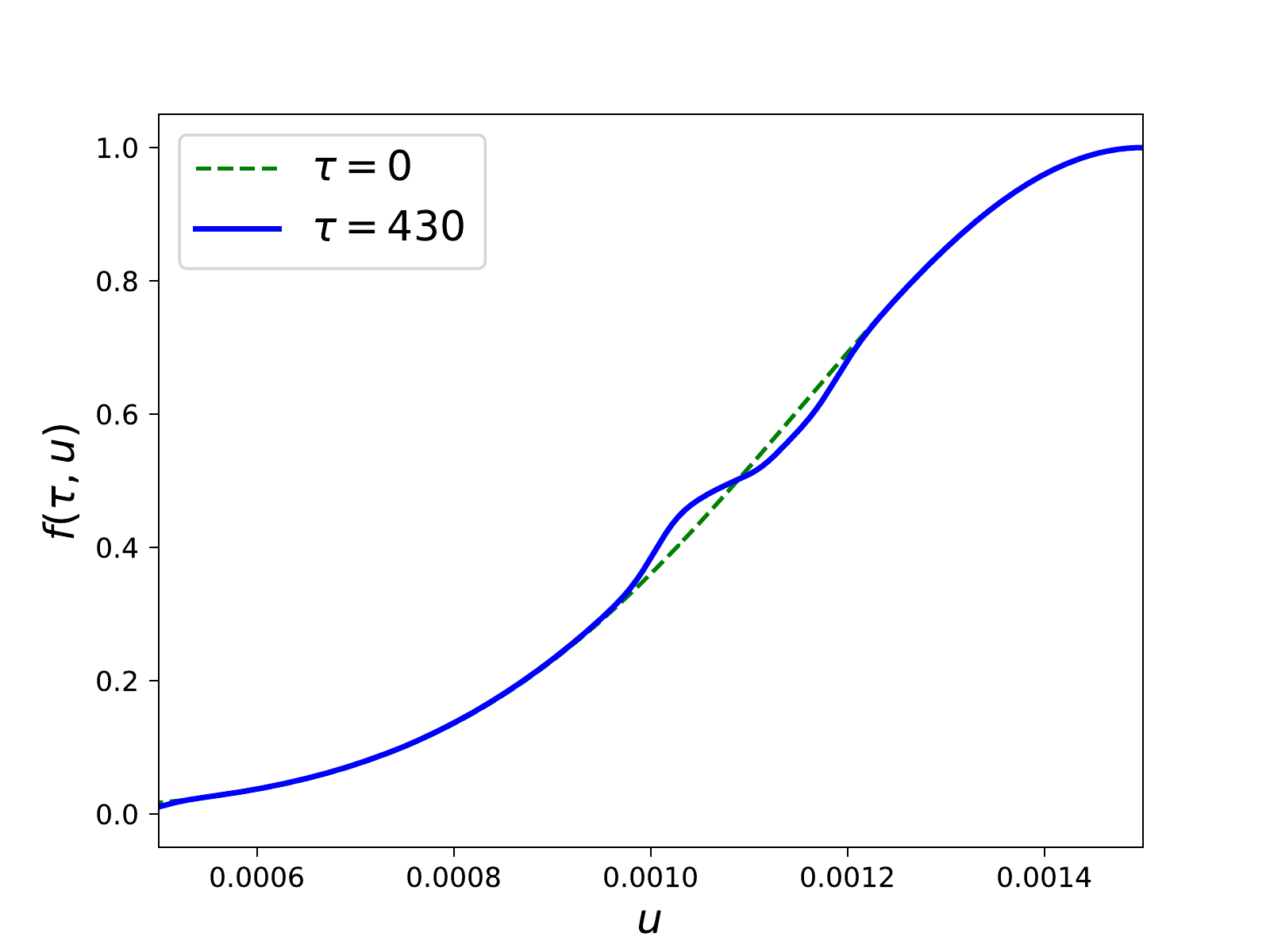}
\caption{(Color online) Upper panel: temporal evolution of the electrostatic potential amplitude integrated from normalized \erefs{eq.hamiltonian.sys} for $\eta=0.0035$ and $\ell_r=912$. The vertical dashed red line represents the saturation time (as indicated in the plot). Lower panel: correspondent zoom on the positive slope of the distribution function taken at saturation time $\tau=430$ (blue solid). The green dashed line here corresponds to the initial energetic particle profile $f_{in}$ of \eref{eq.gaussian}.
\label{fig.f0-fsat}}
\end{figure}
We assume that the initial warm beam is initially distributed in the velocity space as
\begin{equation}
\label{eq.gaussian}
f_{in}(u)=f(\tau=0,u)=e^{-0.5(u-a)^2/b^2}\,,
\end{equation}
with $a\simeq 0.0015$ and $b\simeq 0.00035$. The non-linear simulations are run for a total $N=10^{6}$ particles using a Runge-Kutta (4th order) algorithm. The initial conditions for positions $\bar{x}_i$ are set uniformly in $[0,\,2\pi]$, and the modes are initialized with amplitudes $\mathcal{O}(10^{-14})$ (this ensure the initial linear regime).

For a selected case of interest $M=1$ ($\ell_r=1/u_r=912$), the results are shown in \figref{fig.f0-fsat} for the mode evolution (upper panel) and for the distribution function (lower panel). It clearly emerges the initial exponential growth of the mode amplitude (linear phase), predicted by the linear theory, followed by the non-linear saturation when beam particles became trapped inside the potential wells. After the mode saturation occurring at $\tau_S$ (in this case, $\tau_S\simeq 430$) the amplitude fluctuates near a constant value, which results to be closed to the saturated amplitude of the mode (dubbed $\phi(\tau_S)$). At the same time the distribution function flattened near the resonance velocity ($u_r\simeq 0.0011$) (for details, see \citerefs{ncentropy,nceps16,VK12}). 

In particular, in \citeref{ncEPS18}, the relevant scalings of the system have been pointed out and summarized. The saturated field scales quadratically with the observed linear growth rate $\gamma_L$, \ie $|\phi(\tau_S)|\propto \gamma_L^2$, while the flattening with of the distribution profile, namely the non-linear velocity spread $\Delta u_{NL}$, scales linearly as a function the same quantity with the general law $\Delta u_{NL}/u_r\simeq8.5\gamma_L$ (in the following plots, we will indicate such a spread with dashed vertical lines).

\subsection{Analytical solution}
Let us now compare the analytic solution in \eref{eq.mono} with respect to the numerical simulations sketched above. In order to fix the (normalized) $\alpha$ parameter of \eref{alphapar}, we consider a constant values of the mode taken as the time average of the field after saturation and, from \figref{fig.f0-fsat}, it clearly turns out how this value is close to $|\phi(\tau_S)|$.
\begin{figure}[ht]
\centering
\includegraphics[width=0.8\columnwidth]{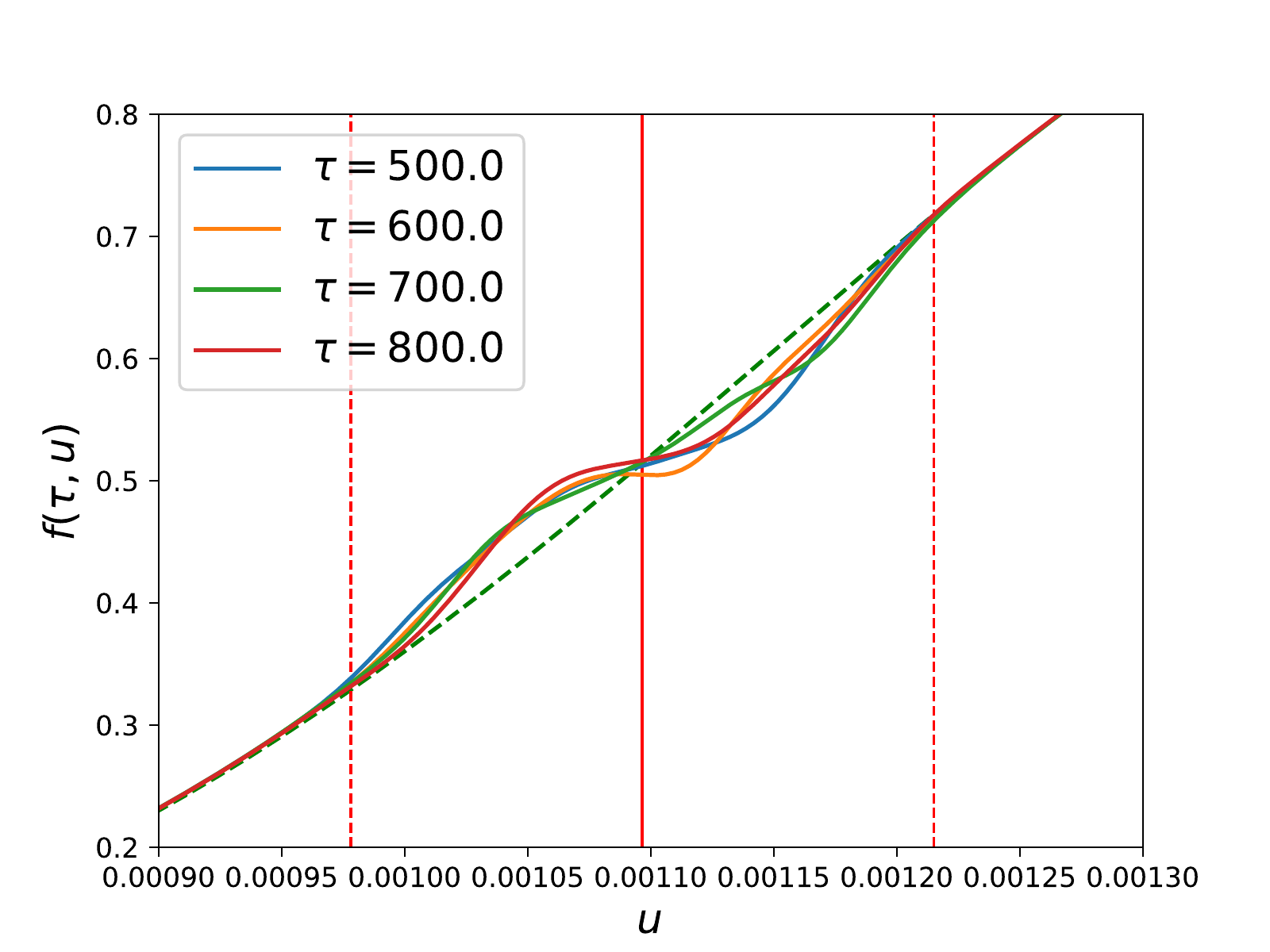}
\caption{(Color online) Evolution of the distribution function evaluated from the numerical simulation of the fully self-consistent $N$-body system \erefs{eq.hamiltonian.sys}, for the selected case $M=1$, $\ell_r=912$ and $\eta=0.0035$ (same as \figref{fig.f0-fsat}). The color scheme is indicated in the plot for different times, while the green dashed line corresponds to the initial $f_{in}$ of \eref{eq.gaussian}. Solid vertical red line denotes $u_r$ and dashed vertical red lines indicate the flattening width predicted by $u_r\pm\Delta u_{NL}$ (the value of $u_r$ is constant in time, due to the absence of frequency sweeping).\label{fig.AKvsNB.NB}}
\end{figure}
It is important to stress that $\alpha$ is thus related to the non-linear velocity spread introduced above.

Introducing the same scaled variables defined in the previous subsection, we have to assign the initial condition for \eref{eq.mono}. Such an initial profile can not concern the linear growth of the mode, since \eref{eq.mono} has been derived assuming a constant mode amplitude. Therefore, we should assign the initial distribution function taking the one obtained in the simulations at $\tau_S$, \ie $f_S(u)=f(\tau_S,u)$, when the mode amplitude almost froze in. This setup was done by fitting numerical data via the function
\begin{align}
\label{eq.fsat}
f_S(u)=f_{in}-[A(C-u)+B]e^{-0.5(u-C)^4/D^4}\,,
\end{align}
($A=794.5$, $B=0.0028$, $C=0.0011$ and $D=0.000075$).

\begin{figure}[ht]
\centering
\includegraphics[width=0.8\columnwidth]{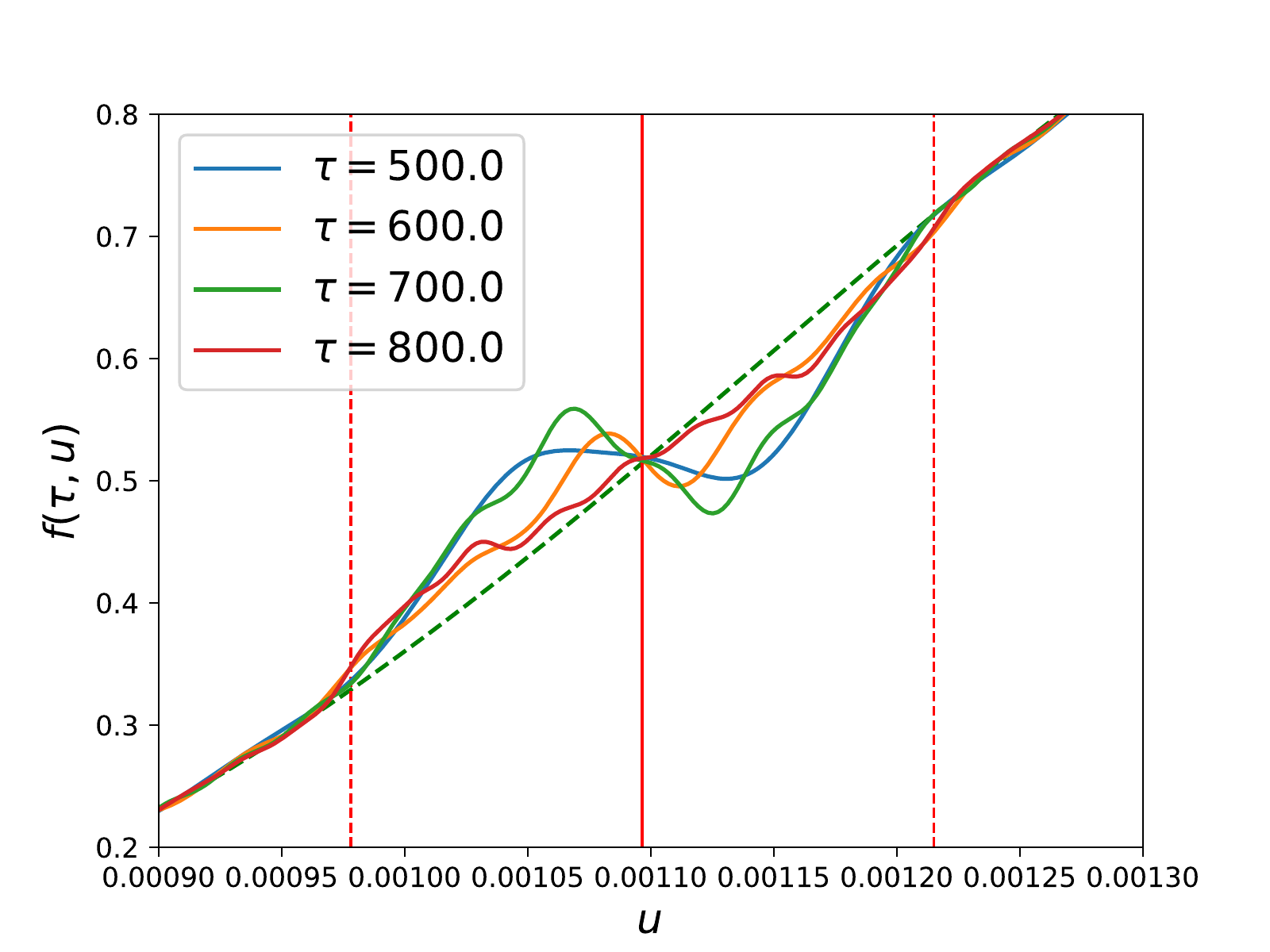}
\caption{(Color online) Evolution of the distribution profile from analytical solution of normalized \eref{eq.mono}. Color scheme and other notations and definitions are from \figref{fig.AKvsNB.NB}. \label{fig.AKvsNB.AK}}
\end{figure}

Using \eref{eq.beta}, we evaluate $\beta_n$ up to $n=58$. Such a limit is imposed by the estimated errors on numerical integrations carried out by means of QUADPACK library \cite{QUADPACK}. The evolution of the distribution function at four instants after the saturation is plotted in \figref{fig.AKvsNB.NB} as obtained by numerical simulations via the $N$-body code, and in \figref{fig.AKvsNB.AK} as predicted by the solution in \eref{eq.mono}.

It clearly emerges that the analytic model and the simulations agree about the position of resonance and its non-linear velocity spread. However, as already discussed, such a spread is fixed by $\alpha$, \ie by the post-saturation averaged mode amplitude taken from the simulations, and therefore this cannot be regarded as an independent prevision. Two main features about \figref{fig.AKvsNB.AK} stand out: the reversal of the slope and the bumpiness of the distribution profiles. The inverse gradient feature outlines that beam electrons have an exceeding loss of velocity with respect to the real system, reliably due to an over energy supply for maintaining constant the imposed filed amplitude. The corrugation of the profile could seem an artifact caused by the truncation of a series expansion, \emph{e.g.} the Gibbs phenomenon for the Fourier series \cite{gibbs}, in this case an expansion in Hermite polynomials \cite{bilodeau}. We will return later on this issue, for now the basic question remains whether the disagreement found is mainly caused by the breaking of self-consistency or by the Dyson-like structure at the base of the model, \ie the choice to resum only the most secular term at each order.

\section{Dyson model with external field}
In this Section, we present a Dyson-like equation for the distribution function which can be numerical solved with an arbitrary assigned field. The aim is to study what happens if the self-consistency is restored, albeit in a rather artificial way. In fact, we will use an external field for the Dyson equation but its form comes out from the purely self-consistent $N$-body code.

Starting from the Vlasov equation in the Fourier space, we consider the dynamics of homogeneous and inhomogeneous terms separately:
\begin{subequations}
\label{eq.vlasov.fourier}
\begin{align}
\label{eq.vlasov.f0}
&\partial_tf_0(t,v)=\frac{e}{m}\sum_{k>0}E_k^*(t)\partial_vf_k(t,v)+c.c.\,,\\
\label{eq.vlasov.fk}
&\partial_tf_k(t,v)=-ikvf_k(t,v)+\frac{e}{m}E_k(t)\partial_vf_0(t,v)\,.
\end{align}
\end{subequations}
In \eref{eq.vlasov.fk}, we neglect all terms $E_{k-q}f_q$ for $q\neq 0$ because of $2$nd order in the field (as done in the previous Section, we consider an homogeneous initial condition, therefore $f_0(t,v)\sim f(v)$ is dominant). The solution of \eref{eq.vlasov.fk} with zero initial condition results in
\begin{equation}
\label{eq.fk}
f_k(t,v)=\frac{e}{m}\int_0^t dt'\,e^{ikv(t-t')}E_k(t')\partial_vf_0(t',v)\,,
\end{equation}
which replaced in \eref{eq.vlasov.f0} provides
\begin{subequations}
\label{eq.dysonext}
\begin{align}
&\label{eq.f0}
\partial_tf_0(t,v)=\left(\frac{e}{m}\right)^2\partial_v\sum_{k>0}[E_k^*(t)g_k(t,v)+c.c.]\,,\\
\label{eq.gq}
&\partial_tg_k(t,v)=-iqvg_k(t,v)+E_k(t)\partial_vf_0(t,v)\,,
\end{align}
\end{subequations}
where we introduced the auxiliary function $g_k$ defined by the second of the equations above.

\subsection{Constant amplitude field closure}
\begin{figure}[ht]
\centering
\includegraphics[width=0.8\columnwidth]{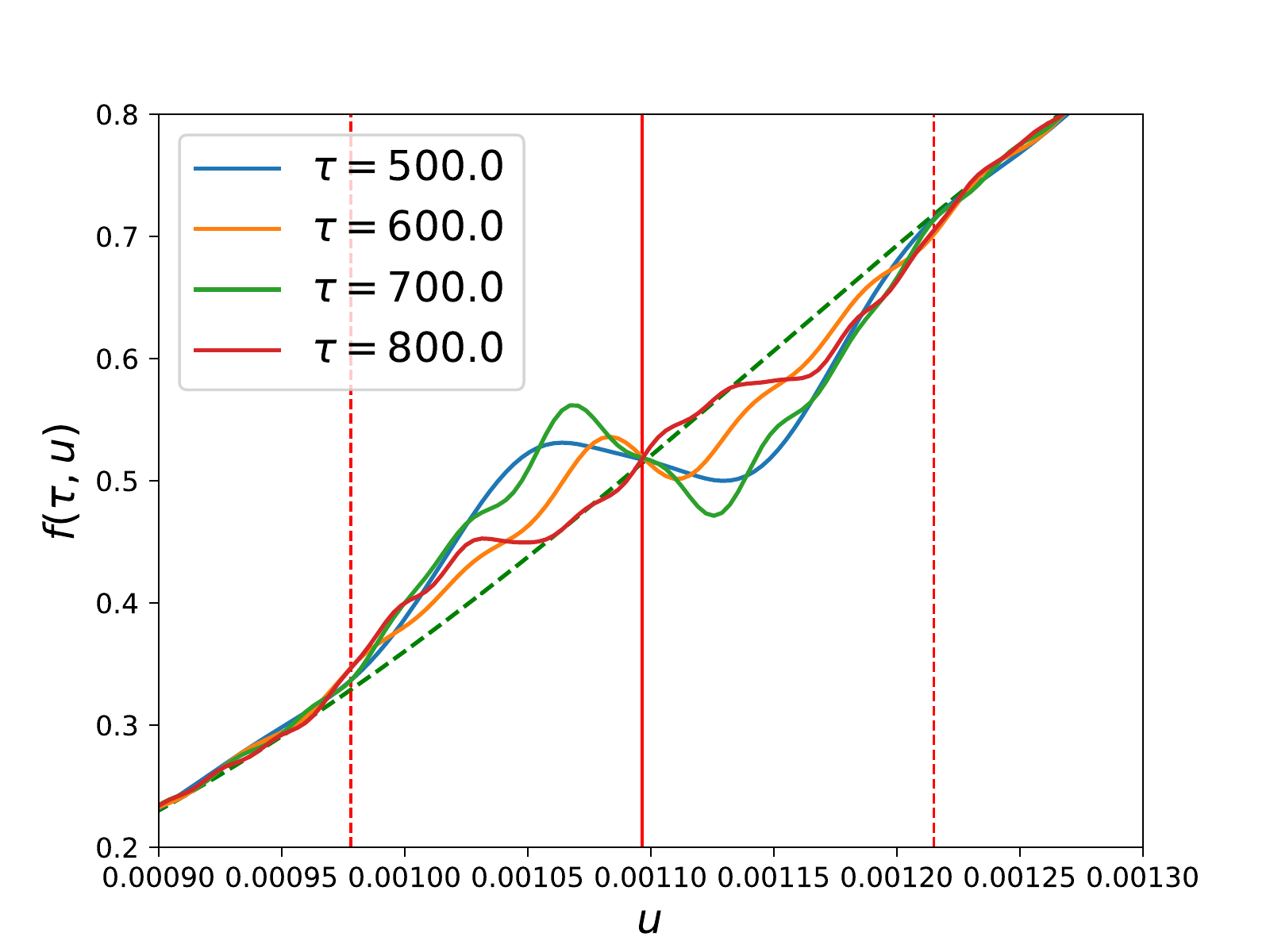}
\caption{(Color online) Distribution function evolved from \erefs{eq.dysonext} by assuming a constant amplitude field (same setup of \figref{fig.AKvsNB.AK}). Color scheme and other notations and definitions are from \figref{fig.AKvsNB.NB}.}
\label{fig.DE.constant}
\end{figure}

Let us now analyze the behavior in time of the distribution function when the external field is assigned with a constant amplitude, \ie $|E_k(\tau)|=|E_k^{(0)}|$. The aim of this analysis is the comparison with the analytical solution in \eref{eq.mono}, in order to characterize the real source of the evolved profile corrugations described above. We thus numerically integrate \erefs{eq.dysonext} assuming a single mode with constant amplitude (same average of the analytical case) and using the same set of initial conditions (\ie the post-saturation distribution). Adopting the same normalization of the considered variables, the results of the numerical integration of \erefs{eq.dysonext} are plotted in \figref{fig.DE.constant}.

From the comparison with \figref{fig.AKvsNB.AK}, it is evident how \erefs{eq.dysonext} with the constant amplitude field closure significantly reproduce the analytical results based on the expansion in the eigenfunctions (truncated at a given order). We argue from this that not only the two aforementioned models have the same Dyson-like structure, but also that the bumpiness already outlined is not an artifact introduced by the expansion truncation. The former statement means also that the \emph{ansatz} at the base of the two models are equivalent, \ie a partial resummation with only the one-loop diagram. This has the same consequences to neglect the terms $E_{k-q}f_q$, which couple different distribution Fourier components.

\subsection{Implication of the self-consistency}
We can now integrate \erefs{eq.dysonext}, by assigning the self-consistent field evolved from \erefs{eq.hamiltonian.sys}.
\begin{figure}[ht]
\centering
\includegraphics[width=0.8\columnwidth]{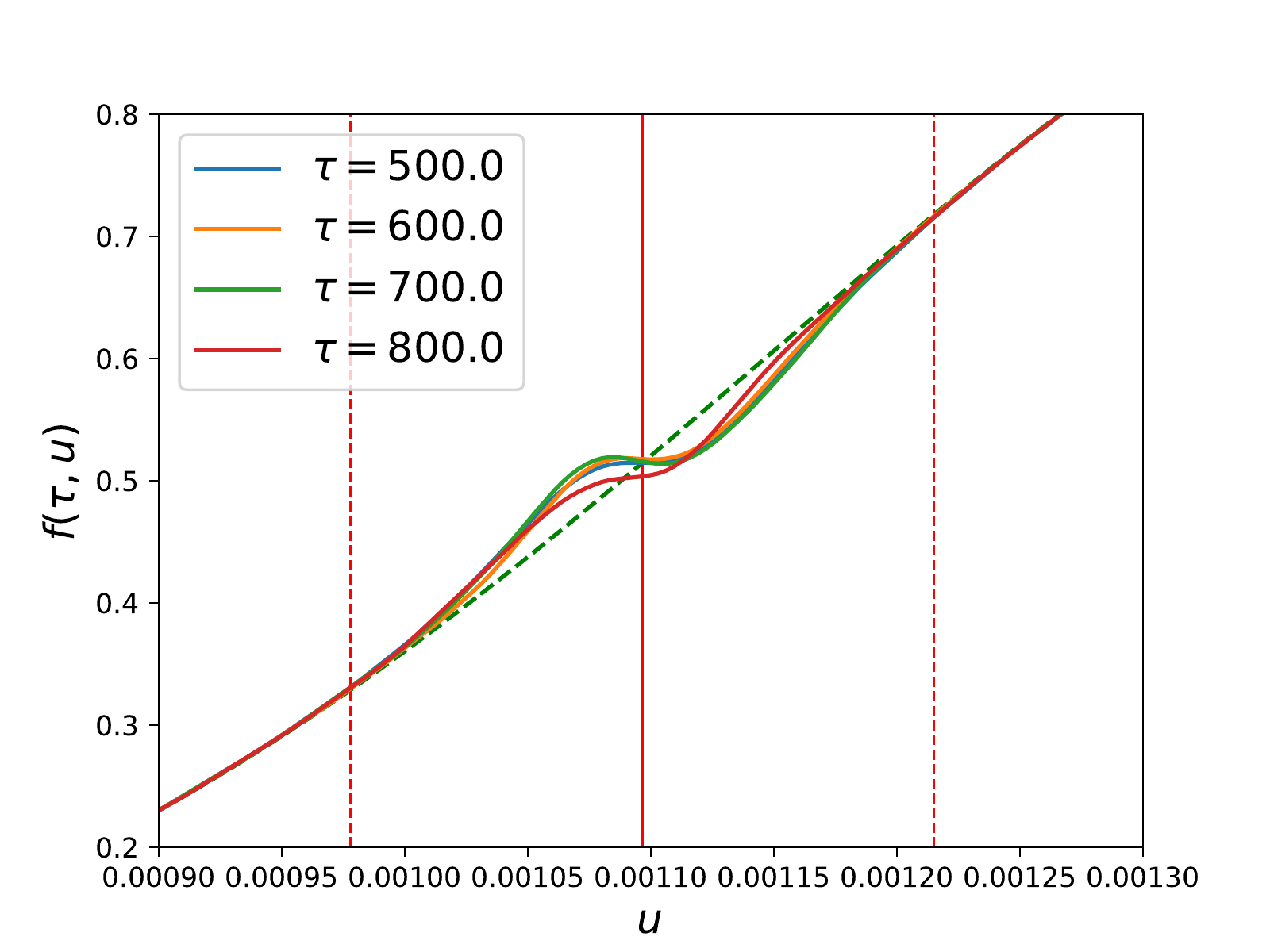}
\caption{(Color online) Evolution of the distribution profile from the numerical integration of \erefs{eq.dysonext}, with assigned self-consistent mode generated by\eref{eq.hamiltonian.sys}. Color scheme and other notations and definitions are from \figref{fig.AKvsNB.NB}.}
\label{fig.DE}
\end{figure}
The results are shown in \figref{fig.DE} and have to be compared with respect to \figref{fig.AKvsNB.NB}. Such a comparison sheds light on the addressed question: the main shortcoming of the analytical solution presented in \citeref{AK66} is the self-consistency breaking, rather than the Dyson-like procedure itself. Actually, by artificially restoring the self-consistency, \ie considering the external field with its proper evolution, the qualitative and quantitative agreements with respect to the $N$-body dynamics is remarkably improved. In fact, the slope inversion is quite suppressed and corrugations disappear.

The main merit of the present analysis is to outline how the Dyson-like procedure of summing the most important diagram only does not prevent a satisfactory description of the Vlasov-Poisson dynamics, provided that the self-consistency is not violated.

\section{Concluding remarks}

We have studied the predictivity of the analytical model proposed in \citeref{AK66}, to describe the beam-plasma instability when the electric field saturates and it is assumed constant. Our investigation is based on the comparison of the analytical Dyson approach with respect to a simulation experiment, characterized in terms of the Hamiltonian description discussed in \citerefs{OWM71,CFMZJPP}.

As a first step, we have recognized that the prediction of the Dyson equation, in correspondence to a fixed amplitude of the electric field, can describe only very qualitative features of the real dynamics: corrugations of the distribution profile and inversion of the velocity gradients take place. Then, we have restated the problem without using the Hermite polynomial expansion, but still retaining fixed the electric field amplitude. We have shown that the result is qualitatively similar to the previous case, \ie we gain nothing in the capability to predict the simulation experiment. Finally, we have used this scheme by inserting in the Dyson equation the electric field evolution obtained by the simulation experiment, so showing that the resulting distribution function acquires a more realistic profile. 

We argue that the present study can be regarded as paradigmatic of the shortcomings of breaking down the self-consistency of the field amplitude and the particle distribution function profile even in more general context. Clearly, we can also upgrade the present model by accounting for higher order contributions in the dyagrammatic schematization proposed in \citeref{AK66}, which could be responsible for non-diagonal couplings of the monochromatic modes.

\acknowledgments
{\footnotesize This work has been partly carried out within the framework of the EUROfusion Consortium [Enabling Research Projects: NAT (AWP17-ENR-MFE-MPG-01), MET (CfP-AWP19-ENR-01-ENEA-05)] and has received funding from the Euratom research and training programme 2014-2018 and 2019-2020 under grant agreement No 633053. The views and opinions expressed herein do not necessarily reflect those of the European Commission.}


\end{document}